# User Acceptance of Usable Blockchain-Based Research Data Sharing System: An Extended TAM-Based Study

Ajay Kumar Shrestha
Department of Computer Science
University of Saskatchewan
Saskatoon, Saskatchewan, Canada
ajay.shrestha@usask.ca

Julita Vassileva
Department of Computer Science
University of Saskatchewan
Saskatoon, Saskatchewan, Canada
julita.vassileva @usask.ca

*Abstract*—Blockchain technology has evolved as a promising means to transform data management models in many domains including healthcare, agricultural research, tourism domains etc. In the research community, a usable blockchain-based system can allow users to create a proof of ownership and provenance of the research work, share research data without losing control and ownership of it, provide incentives for sharing and give users full transparency and control over who access their data, when and for what purpose. The initial adoption of such blockchain-based systems is necessary for continued use of the services, but their user acceptance behavioral model has not been well investigated in the literature. In this paper, we take the Technology Acceptance Model (TAM) as a foundation and extend the external constructs to uncover how the perceived ease of use, perceived usability, quality of the system and perceived enjoyment influence the intention to use the blockchain-based system. We based our study on user evaluation of a prototype of a blockchain-based research data sharing framework using a TAM validated questionnaire. Our results show that, overall, all the individual constructs of the behavior model significantly influence the intention to use the system while their collective effect is found to be insignificant. The quality of the system and the perceived enjoyment have stronger influence on the perceived usefulness. However, the effect of perceived ease of use on the perceived usefulness is not supported. Finally, we discuss the implications of our findings.

## I. INTRODUCTION

In the research community, data sharing practices are much needed to maximize the knowledge gains from the research efforts, reduce duplicative trials and accelerate discovery. In medicine and healthcare, both personalized patient care and medical research can benefit from ethical and privacy-preserving sharing of patient data and data from clinical trials [1]. A flexible mechanism for obtaining and renewing consent for data use and sharing is required that provides appropriate and meaningful incentives to capitalize from data sharing and ensures transparency for users to be aware of which of their data has been accessed, by whom, for what purpose and under what conditions.

Currently, blockchain technology and smart contracts have evolved as a promising means to support immutable and trusted in various use fields including research community [2], healthcare [3], agricultural [4], tourism domains [5] etc. Initially, blockchain was used only to implement virtual currencies [6], but the applications of the blockchain technology have since quickly evolved to a wide variety of use cases [7]. Smart contracts committed on the blockchain can encode allowed purposes of data use, allowed software apps, people or businesses who can access the data, time limitations, price for access, etc. in various usable cases. The key idea is that the ledger in the blockchain is neither stored in a centralized location nor managed by any single entity; multiple distributed parties come to a consensus, which is committed into the ledger and thereafter can be accessed by anyone. Computationally, it is impracticable for any corrupted node (unless the number of such nodes is higher majority consensus) to go back and alter the history. There is no single point of failure in blockchain because the redundancy of the system ensures many backups, and the lack of a central storage place ensures there is no one target for hackers [8]. Therefore, blockchain provides a new type of platform that is useful for sharing research data by providing solutions to the problems of privacy of user data and compliance to ethics standards and user consent agreements, as well as researcher control and incentives for sharing. The most important criticisms to blockchain-based approaches to date relate to their performance and scalability; yet the rapid development of the technology allows, through thoughtful combinations of blockchains to achieve acceptable performance. A harder problem emerges, related to the user acceptance of blockchain technology in non-currency related application domains.

For example, due to the lack of familiarity with blockchain, it is not clear if researchers would be receptive to using blockchain technology in regulating access and sharing of research data. It is therefore important to study the user acceptance of blockchain-based applications for example if users understand the smart contracts and blockchain technologies and if they can competently share research data. Many studies have evaluated the performance of blockchain-based systems [5], [9], [10], [11]. However, to our best knowledge, no studies has focused on user acceptance of blockchain-based system. To bridge this gap and advance research in blockchains- and smart contracts-based systems, we adopted the extended Technology Acceptance Model (TAM) to examine the indicators that affect the user's acceptance of the system. We based our study on user evaluation of a prototype of a blockchain-based research data-sharing framework [2] with the TAM validated tool deployed as a research instrument to collect data from 20 participants. We chose to investigate the influence of the perceived ease of use, perceived usefulness, perceived enjoyment and quality of the system, on the participants' intention to use the system. We also analysed the influence of the perceived ease of use, perceived enjoyment and quality of system on perceived usefulness.

The results of our investigation show a stronger influence of quality of system ($\beta = 0.83$, $p < 0.0001$) and perceived enjoyment ($\beta = 0.75$, $p < 0.0001$) on intention to use for the blockchain-based research data sharing system while perceived usefulness (β = 0.5, p < 0.01) and perceived ease of use (β = 0.56, p < 0.05) have moderate and weaker effects respectively. Moreover, our results show that combined effect of all four antecedents, perceived ease of use (β = -0.045, p > 0.05), perceived usefulness (β = 0.053, p > 0.05), quality of system (β = 0.364, p > 0.05) and perceived enjoyment (β = 0.48, p > 0.05) on intention to use ($R^2$ = 0.75) is found to be insignificant. Specifically, our results show that effect of perceived ease of use (β = 0.45, p > 0.05) on perceived usefulness is unaccepted. However, the quality of system (β = 0.99, p < 0.001) and perceived enjoyment (β = 0.75, p < 0.01) have stronger influence on perceived usefulness.

In the next section, we present some background and related works on blockchains- and smart contracts-based research data sharing system and extended TAM model. Sections III presents our research method with research questions, measurement instruments and the demographics of participants in our survey. The descriptive statistics, measurement models, structural models and brief analysis of the results are presented and discussed in section IV. Finally, section V concludes the paper with future research directions.

## II. BACKGROUND AND RELATED WORK

### A. Blockchain-based research data sharing system

Most researchers, on an individual level, may feel reluctant to share their research data; however, they appreciate the overall benefits of data sharing, which was also concluded from the qualitative interviews-based study conducted in [12], [13]. Those studies recognized six different ways of data sharing: private management sharing, peer exchange, community sharing, collaborative sharing, sharing for transparent government and public sharing.

There are proposals in the literature (most prominently, [14]) to use blockchain as access control platform to ensure the privacy of data. In [15], the authors mentioned some of the technical challenges present in the proposals adopting blockchain as part of their solutions. We proposed [2] a usable blockchain-based model for data sharing in the scientific research domain which addresses these challenges. Two different blockchains are used, providing accountability of access and incentives for sharing, maintaining the complete and updated information, and a verifiable record of the provenance, including all accesses/sharing/usages of the data. This model is the basis for the user-acceptance study that we present in this paper, but since the technical detail is not essential for the study, the interested reader is referred to [2] for a full description.

### B. Extended TAM Model

The Technology Acceptance Model (TAM) as shown in Figure 1 [16], [17] was introduced on the basis of Theory of Reasoned Action (TRA) [18], which claims that the behavioral intention is a strong indicator of actual behavior.

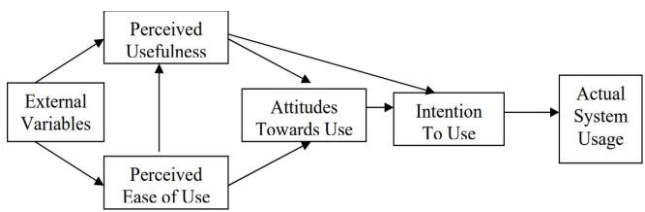

Fig. 1. A classical TAM model

TABLE I. CONSTRUCTS AND DEFINITION [17], [19], [20]

| Construct | Definition |
|---|---|
| Perceived Ease of Use (PEOU) | It is the degree to which a person believes that using a particular system would be free of effort. |
| Perceived Usefulness (PU) | It is the degree to which a person believes that using a particular system would enhance his or her job performance. |
| Quality of System (QOS) | It is the degree to which a person is pleased, hence reducing users' psychological objection to the system or the loss of volition. |
| Perceived Enjoyment (PEnj) | It is the degree to which the use of technology is seen to be enjoyable. |
| Intention to use (ITU) | It is the degree to which a person has behavioral intention to adopt the technology. |

The TAM model has been used as a conceptual framework in many literatures to study the potential users' behavioral intention to use a particular technology. The *behavioral intention* is define as "the degree to which a person has formulated conscious plans to perform or not perform some specified future behavior" [21], which is therefore in line with the TRA. The classical TAM focuses on using technology, where perceived ease of use (PEOU) and perceived usefulness (PU) are two factors or antecedent to influence user acceptance behavior.

It hypothesizes that the actual use of system is determined by behavioral intention to use, which is in turn influenced by user's attitude toward using the system and perceived usefulness and perceived ease of use of the system as represented in Figure 1. However, many researchers often extend TAM by adding external constructs depending upon the contexts due to the limitation of classical TAM from the fact that many important factors are not included in the model [22]. Perceived enjoyment [19], quality of system [20], trust [23], behavioral control [24] etc. are some of the constructs that have been added as influential variables to user acceptance of the information technology.

TAM is widely used to understand how users come to accept and use information technology. However, there is no literature on TAM in the context of blockchains and smart contracts-based applications, indicating a significant gap in knowledge. To fill this gap our research applies the extended TAM to distributed ledger technologies.

Table I presents the definition of different study's constructs for the extended TAM that we used in our research. The corresponding extended TAM is shown in Figure 2. Based on the literature [17], [19], [20] about user study, we conducted a similar study to investigate the user perception on the acceptance of the blockchain-based applications.

## III. METHODOLOGY

In this section, we present our research hypotheses, research questions, measurement instruments and the demographics of participants.

### A. Research Hypotheses

We based our study on user evaluation of prototypes of the blockchain-based research data-sharing framework with the questionnaire deployed as a research instrument to collect data. We set several hypotheses, in our context based on the literature review to investigate the constructs as given in Table I, which are as follows:

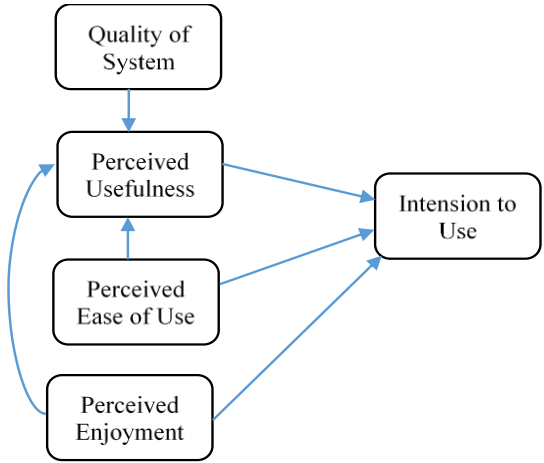

Fig. 2. An Extended TAM model for our study

- H1: The *perceived ease of use* will significantly influence the *perceived usefulness* of the blockchain-based research data-sharing framework.
- H2: The *perceived enjoyment* will significantly influence the *perceived usefulness* of the blockchain-based research data-sharing framework.
- H3: The *quality of system* will significantly influence the *perceived usefulness* of the blockchain-based research data-sharing framework.
- H4: The *perceived ease of use* will significantly influence the *intention to use* the blockchain-based research data-sharing framework.
- H5: The *perceived usefulness* will significantly influence the *intention to use* the blockchain-based research data-sharing framework.
- H6: The *quality of system* will significantly influence the *intention to use* the blockchain-based research data-sharing framework.
- H7: The *perceived enjoyment* will significantly influence the *intention to use* the blockchain-based research data-sharing framework.
- H8: The combined effect of *perceived ease of use*, *perceived usefulness, quality of the system and* perceived *enjoyment* will significantly influence the *intention to use* the blockchain-based research data-sharing framework.

### B. Research Design

The study was approved by the Behavioral Research Ethics Board of the authors' university. To contextualize the extended TAM tool, we provided participants at the beginning with a brief description of our blockchain-based research data-sharing framework proposed in [2]. Thereafter, we presented the participants with an online survey through SurveyMonkey. The survey instrument is based on constructs validated in [17], [19], [20] and adapted to the context of our study. The instrument consists of six items for perceived ease of use, six items for perceived usefulness, four items for quality of system, three items for perceived enjoyment and four items for intention to use. The respective items (questions) in the constructs are shown in Table II. We also added a question to rate the user's confidence level while getting their input for every construct.

TABLE II. CONSTRUCTS AND ITEMS [17], [19], [20]

| Construct | Items |
|---|---|
| Perceived Ease of Use (PEOU) | peou1 - Learning to operate this system would be easy for me. |
| | peou2 - I would find it easy to get this system to do what I want it to do. |
| | peou3 - My interaction with this system would be clear and understandable. |
| | peou4 - I would find this system to be flexible to interact with. |
| | peou5 - It would be easy for me to become skillful at using this system. |
| | peou6 - I would find this system easy to use. |
| | How confident are you in the ratings made on this page? |
| Perceived Usefulness (PU) | pu1 - Using this system would enable me to accomplish data sharing tasks more quickly. |
| | pu2 - Using this system would improve my performance with regard to sharing research data. |
| | pu3 - Using this system would increase my productivity. |
| | pu4 - Using this system would increase my effectiveness. |
| | pu5 - Using this system would make it easier to share the data. |
| | How confident are you in the ratings made on this page? |
| Quality of System (QOS) | qos1 - I would be satisfied with the research paper sharing methodology of this system. |
| | qos2 - I would be satisfied with the feature of creating proof of the existence of the research work (ownership). |
| | qos3 - I would be satisfied with the feature of allowing users to set permissions for the way to share their data. |
| | How confident are you in the ratings made on this page? |
| Perceived Enjoyment (PEnj) | penj1 - I would be satisfied to use this system to share research data |
| | penj2 - I would like to use this system to share research data. |
| | How confident are you in the ratings made on this page? |
| Intention to use (ITU) | itu1 - I believe it is worthwhile to use this system to share research data. |
| | itu2 - I will use this system to share research data. |
| | itu3 - I intend to use this system to share research data in the future. |
| | How confident are you in the ratings made on this page? |

TABLE III. PARTICIPANTS' DEMOGRAPHICS

| Respondents' characteristics [(Female, male) = (45%, 55%)] | Criteria | Percentage |
|---|---|---|
| Age | 18 to 24 | 9.09% |
| | 25 to 34 | 72.73% |
| | 35 to 44 | 18.18% |
| Highest education level | Grad-High school | 4.55% |
| | Bachelors | 22.73% |
| | Masters | 54.55% |
| | PhD | 18.18% |
| Current occupation | Student | 59.09% |
| | Researcher | 31.82% |
| | Faculty | 4.55% |
| | Other | 4.55% |
| Ever served as a reviewer | Yes | 42.86% |
| | No | 57.14% |
| Familiar with blockchain technologies and smart contracts | Extremely familiar | 13.64% |
| | Very familiar | 27.27% |
| | Somewhat familiar | 27.27% |
| | Not so familiar | 31.82% |
| Familiar with research/ social networks (e.g. ResearchGate, Mendeley, ORCID) | Extremely familiar | 9.52% |
| | Very familiar | 42.86% |
| | Somewhat familiar | 47.62% |

We measured the responses to the items on a 7-scale Likert scale from 1 = extremely unlikely to 7 = extremely likely. A total of 22 participants took part in the study, but upon data cleaning, 20 were left for analysis. We recruited participants from Academia, who had some research experience. Specifically, around 47% of participants were somewhat familiar, and 53% were highly familiar with other research content sharing social networks such as ResearchGate, Mendeley or Orchid. Table III highlights the demographics of the participants.

## IV. RESULT

In this section, we first present and briefly analyze the collected data with the descriptive statistic. Then, we present our results of the structural equation model (SEM), which includes the measurement models (internal consistency, composite reliability, average variance extracted, KMO and Bartlett's test of sphericity) and structural models (exploratory factor analysis, regression analysis) and brief analysis of the results. For the second part, we started by fitting the measurement models to the data, and later we tested the underlying structural models. The calculations of descriptive statistics in this study were carried out using MS Excel and IBM SPSS Statistics 25.

### A. Descriptive Statistic

Since we measured the responses to the items on a 7-scale Likert scale, we categorized the scale in terms of percentage value to analyze the score for each item and overall impression of the construct. Table IV provides the category of a percentage value for seven different levels of Likert scale. Table V to Table IX summarizes data collected for all the items in perceived ease of use, perceived usefulness, quality of system, perceived enjoyment and intention to use constructs of our model respectively.

The obtained scores for different selected constructs indicate that user perceptions on the benefits of using proposed should be maintained or enhanced by making improvements in order to achieve higher level of score category. The preliminary descriptive statistic of the obtained data shows that all of the constructs provide a significant impression in the context of user acceptance of the usable blockchain-based research data sharing prototype.

TABLE IV. CATEGORIZATION FOR PERCENTAGE VALUE

| Value | Category |
|---|---|
| 85.71 < x ≤ 100 | Extremely High |
| 71.43 < x ≤ 85.71 | Quite High |
| 57.14 < x ≤ 71.43 | Slightly High |
| 42.86 < x ≤ 57.14 | Neither |
| 28.57 < x ≤ 42.86 | Slightly Low |
| 14.29 < x ≤ 28.57 | Quite Low |
| 0 < x ≤ 14.29 | Extremely Low |

TABLE V. ANALYSIS OF PERCEIVED EASE OF USE (PEOU)

| Indicators | Score | Percentage | Std. Deviation |
|---|---|---|---|
| Ease of Learning | 6 | 85.72 | 0.726 |
| Controllable | 5.65 | 80.72 | 1.04 |
| Understandable | 5.55 | 79.29 | 0.945 |
| Flexible | 5.7 | 81.43 | 1.129 |
| Effort to Skillful | 5.75 | 82.15 | 0.911 |
| Easy to Use | 5.8 | 82.86 | 1.057 |
| **Total Average** | 5.742 | 82.03 | |
| **Confidence in the rating** | 5.7 | 81.43 | |
| **Category** | | Quite High | |

TABLE VI. ANALYSIS OF PERCEIVED USEFULNESS (PU)

| Indicators | Score | Percentage | Std. Deviation |
|---|---|---|---|
| Work More Quickly | **5.65** | 80.72 | 0.934 |
| Job Performance | **5.3** | 75.72 | 1.261 |
| Increase Productivity | **5.1** | 72.86 | 1.411 |
| Effectiveness | 4.95 | 70.72 | 1.539 |
| Makes Job Easier | 5.95 | 85 | 0.999 |
| Useful | 6.05 | 86.43 | 1.191 |
| **Total Average** | 5.5 | 78.58 | |
| **Confidence in the rating** | | 82.86 | |
| **Category** | | Quite High | |

TABLE VII. ANALYSIS OF QUALITY OF SYSTEM (QOS)

| Indicators | Score | Percentage | Std. Deviation |
|---|---|---|---|
| Satisfy with research file sharing method | 5.8 | 82.86 | 0.834 |
| Satisfy with retaining ownership | 5.9 | 84.29 | 0.789 |
| Satisfy with setting permission for data sharing | 6.05 | 86.43 | 0.888 |
| Satisfy with receiving incentives for data sharing | 5.9 | 84.29 | 0.912 |
| **Total Average** | 5.913 | 84.48 | |
| **Confidence in the rating** | | 85.72 | |
| **Category** | | Quite High | |

TABLE VIII. ANALYSIS OF ENJOYMENT (ENJ)

| Indicators | Score | Percentage | Std. Deviation |
|---|---|---|---|
| Satisfy to use the system | 5.8 | 82.86 | 0.895 |
| Use the system | 5.85 | 83.58 | 0.989 |
| Enjoy using the system whenever needed | 5.7 | 81.43 | 0.865 |
| **Total Average** | 5.784 | 82.63 | |
| **Confidence in the rating** | | 85.72 | |
| **Category** | | Quite High | |

TABLE IX. ANALYSIS OF INTENTION TO USE (ITU)

| Indicators | Score | Percentage | Std. Deviation |
|---|---|---|---|
| Worthwhile to use | 6.15 | 87.86 | 0.813 |
| Use for sharing research data | 5.85 | 83.58 | 0.876 |
| Intend to use for sharing research data in future | 5.75 | 82.15 | 0.911 |
| Necessary to use to share research data | 5.15 | 73.58 | 1.226 |
| **Total Average** | 5.725 | 81.79 | |
| **Confidence in the rating** | | 85.72 | |
| **Category** | | Quite High | |

Figure 3 shows the average results of the constructs which are all in the range 71.43% to 85.71%; therefore, they qualify for the *quite high* category.

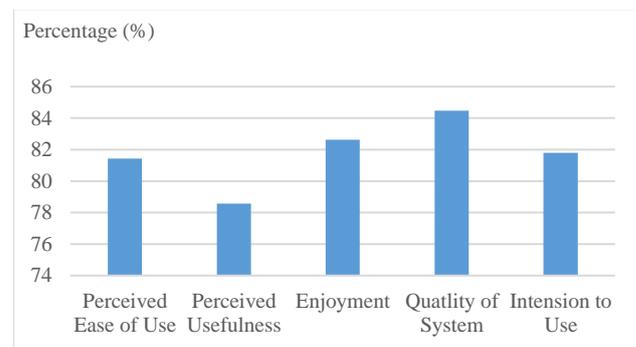

Fig 3. Analysis of all the constructs

TABLE X. EXPLORATORY FACTOR ANALYSIS

| Item | PEOU | PU | QOS | PEnj | ITU | SMC |
|---|---|---|---|---|---|---|
| peou1 | 0.852 | - | - | - | - | 0.722 |
| peou2 | 0.850 | - | - | - | - | 0.722 |
| peou3 | 0.815 | - | - | - | - | 0.664 |
| peou4 | 0.758 | - | - | - | - | 0.574 |
| peou5 | 0.734 | - | - | - | - | 0.538 |
| peou6 | 0.707 | - | - | - | - | 0.499 |
| pu1 | - | 0.694 | - | - | - | 0.481 |
| pu2 | - | 0.892 | - | - | - | 0.795 |
| pu3 | - | 0.786 | - | - | - | 0.617 |
| pu4 | - | 0.836 | - | - | - | 0.698 |
| pu5 | - | 0.752 | - | - | - | 0.565 |
| pu6 | - | 0.876 | - | - | - | 0.767 |
| qos1 | - | - | 0.891 | - | - | 0.793 |
| qos2 | - | - | 0.882 | - | - | 0.777 |
| qos3 | - | - | 0.858 | - | - | 0.736 |
| qos4 | - | - | 0.812 | - | - | 0.659 |
| penj1 | - | - | - | 0.941 | - | 0.885 |
| penj2 | - | - | - | 0.917 | - | 0.840 |
| penj3 | - | - | - | 0.913 | - | 0.833 |
| itu1 | - | - | - | - | 0.879 | 0.772 |
| itu2 | - | - | - | - | 0.868 | 0.753 |
| itu3 | - | - | - | - | 0.799 | 0.638 |
| itu4 | - | - | - | - | 0.663 | 0.439 |

### B. Measurement Models

We checked the measurement model with the exploratory factor analysis by testing the internal data consistency, reliability and validity of the constructs.

*1) Exploratory factor analysis:* Based on the recommendation of Hair et al. [25], factor loadings greater than 0.50 can be considered as significant. We checked the factor loadings in the measurement model to see whether the items in each variable loaded highly on its own construct over the other respective constructs. Table X presents the factor loadings and their corresponding Squared Multiple Correlation (SMC) for our study. All the indicators in the measurement models had a factor loading greater than 0.50.

*2) Convergent Validity*: We observed the convergent validity for each construct measure by calculating Average Variance Extracted (AVE) and Composite Reliability (CR) [25] from the factor loadings (see Table XI). AVE for each construct exceeded the recommended level of 0.50, so over 50% of the variances observed in the items were accounted for by the hypothesized constructs. Similarly, CR should also be above 0.75 to publish result. In our study, CR for each construct was above 0.80.

*3) Reliability of the Measures:* We checked the internal consistency for estimating the reliability of a measure by evaluating the within-scale consistency of the responses to the items of the measure. Since our study has multiple-item measurement instruments, we used Cronbach (Coefficient) Alpha [26] for estimating the internal consistency. "Coefficient Alpha assumes: i) unidimensionality, and that ii) items are equally related to the construct; therefore, interchangeable" [26].

TABLE XI. RELIABILITY ANALYSIS

|  | PEOU | PU | QOS | PEnj | ITU |
|---|---|---|---|---|---|
| Cronbach's α | 0.87 | 0.89 | 0.882 | 0.913 | 0.792 |
| AVE | 0.621 | 0.654 | 0.742 | 0.853 | 0.651 |
| CR | 0.907 | 0.919 | 0.92 | 0.946 | 0.881 |

TABLE XII. DATA SUITABILITY ANALYSIS

|  |  | PEOU | PU | QOS | PEnj | ITU |
|---|---|---|---|---|---|---|
| KMO Measure | | 0.63 | 0.82 | 0.778 | 0.747 | 0.661 |
| Bartlett's Test | χ2 | 59.58 | 69.30 | 39.87 | 36.64 | 29.49 |
|  | df | 15 | 15 | 6 | 3 | 6 |
|  | Sig. | 0 | 0 | 0 | 0 | 0 |

In practice, CR does not assume factor loadings to be the same for all items but take into consideration the varying factor loadings of the items, whereas Alpha assumes factor loadings to be the same for all items. As can be seen in Table XI, the Alpha coefficient for each of the four antecedent construct measure is greater than 0.8 (good) while Alpha for intention to use is greater than 0.75 (acceptable) based on the recommendation of [27]. CR and Alphas are related to each other based on factor loadings as more factor loadings fluctuate among items, the higher the discrepancy between the values of CR and Alpha will be.

*4) KMO and Bartlett's Test:* We performed KMO test for suitability of data for factor analysis based on [28] and found that the KMO measure > 0.5 (acceptable) as can be seen in Table XII. Similarly, based on [28], we then performed Bartlett's Test of sphericity to check the homogeneity of variance for our structural models- ANOVA and regression models. Our result showed that the significance level was smaller than 0.05 as recommended (see Table XII), which suggested the factor analysis would be useful with our data.

### C. Structural Models

We built a global model for the general population in order to begin our Structural Equation Modeling (SEM) analysis [29], as shown in Figure 4. The model is characterized by coefficients of determination ($R^2$'s) and path coefficients (β's). $R^2$ determines the variance of a given construct explained by antecedents, while β captures the strength of the relationship between the selected constructs. The structural model shows different paths linking between perceived ease of use, perceived usefulness, quality of system, perceived enjoyment and intention to use constructs in the context of blockchain-based research data sharing prototype.

TABLE XIII. SEM ANALYSIS

| Structural Path | | β | T Statistics | P-Value | $R^2$ |
|---|---|---|---|---|---|
| PU ← PEOU | | 0.453 | 1.563 | 0.135 | 0.11 |
| PU ← PE | | 0.755 | 3.550 | 0.002 | **0.41** |
| PU ← QOS | | 0.993 | 4.578 | 0.000 | **0.54** |
| ITU ← | PEOU | -0.044 | -0.236 | 0.816 | 0.86 |
|  | PU | 0.053 | 0.341 | 0.737 |  |
|  | QOS | 0.364 | 1.266 | 0.224 |  |
|  | PE | 0.480 | 2.295 | 0.036 |  |
| ITU ← PEOU | | 0.562 | 2.876 | 0.01004 | 0.31 |
| ITU ← PU | | 0.5 | 3.674 | 0.00173 | 0.42 |
| ITU ← QOS | | 0.834 | 5.803 | 0.000 | 0.65 |
| ITU ← PE | | 0.751 | 6.46 | 0.000 | 0.7 |

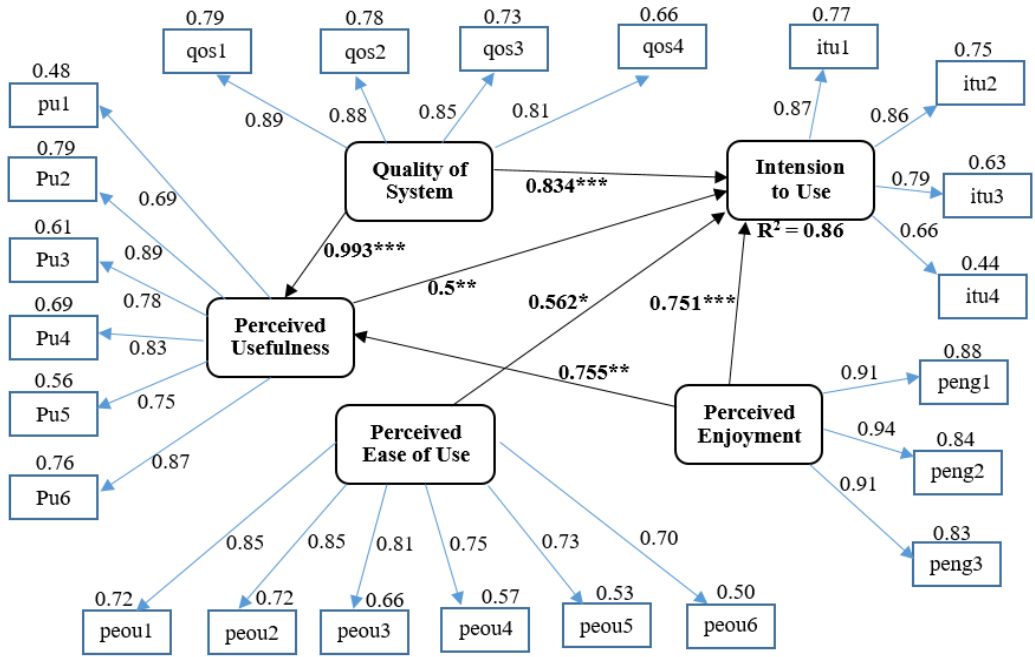

Fig. 3. Structural (Global) model showing test results.
*p < 0.05; **p < 0.01; ***p < 0.001

TABLE XIV. VALIDATION OF STUDY'S HYPOTHESES

| H | Hypothesis | Result |
|---|---|---|
| 1 | The perceived ease of use will significantly influence the perceived usefulness. | × |
| 2 | The perceived enjoyment will significantly influence the perceived usefulness. | √ |
| 3 | The quality of system will significantly influence the perceived. | √ |
| 4 | The perceived ease of use will significantly influence the intention to use. | √ |
| 5 | The perceived usefulness will significantly influence the intention to use. | √ |
| 6 | The quality of system will significantly influence the intention to use. | √ |
| 7 | The perceived enjoyment will significantly influence the intention to use. | √ |
| 8 | The combined effect of perceived ease of use, perceived usefulness, quality of system and perceived enjoyment will significantly influence the intention to use. | × |

√ = True; × = False

Table XIII shows the standardized path coefficient (β), t-statistics, p-value and $R^2$ across selected constructs. According to Chin's guideline [20], a path coefficient should be equal to or greater than 0.2 in order to be considered relevant. Based on [21], we normally refer a model to be statistically somewhat significant (*p) when p-value < 0.05, statistically quite significant (**p) when p-value < 0.01 and statistically highly significant (***p) when p-value < 0.001. In our study, we find that the combined effect of perceived ease of use, perceived usefulness, quality of system, and perceived enjoyment on intention to the blockchain-based research data sharing system are insignificant at p > 0.05. The path coefficients range from -0.044 to 0.480. However, the individual influence of quality of system (β = 0.83, p < 0.001) and perceived enjoyment (β = 0.75, p < 0.001) on intention to use is highly significant while there is a moderate and weaker influence of perceived usefulness (β = 0.5, p < 0.01) and perceived ease of use (β = 0.56, p < 0.05) respectively on intention to use. Hence, hypotheses H4 - H7 are supported, whereas H8 is not supported.

Moreover, we find that perceived ease of use does not relate positively to perceived usefulness (β = 0.453, p > 0.05). However, the quality of system (β = 0.99, p < 0.001) and perceived enjoyment (β = 0.75, p < 0.01) have significant positive effect on perceived usefulness. So, our hypotheses H2 and H3 are also supported, whereas H1 is not supported. Table XIV summarizes the validation of our study's hypotheses.

V. DISCUSSION

We achieved the goal of our research to introduce external constructs, perceived enjoyment and quality of system on the classical TAM in the context of blockchain-based research data sharing system and explore whether users are willing to adopt the system. Our study validates most of the proposed hypotheses. Quality of system is the most significant determinant that influences perceived usefulness and intention to use.

When users receive greater satisfaction with the quality of the blockchain-based system that helps researchers to share their data while maintaining ownership over the data, set permissions to data sharing and receive incentives for sharing the data, the system's perceived usefulness will be higher as well as the user's intention to use it. Furthermore, when users enjoy and get satisfied with the quality of system during their interaction with the prototype system with known benefits for sharing research data, they are quite likely to find the system more useful and extremely likely to adopt the system.

Previous research shows [30], [31], that the UI design is the most significant external construct that affects perceived ease of use, and since our study used a prototype rather than an actual working blockchain based-system, most subjects may have experienced difficulty in relating the actual user-interface. Thus, the effect of easy to use doesn't reflect on the users' belief in finding it to be more useful, which explains our failure to confirm H1 (which is predicted by the classical TAM).

The main limitation of our study is that our findings are based on a small sample size and a prototype-system. Further studies are needed to confirm that the findings generalize for larger population of users in a real system for sharing research data using blockchains. Yet, the methodology for doing a larger study in the context of a real system will be the same.

## VI. CONCLUSION

User studies are much needed to evaluate technological solutions and observe the effects of different variables using theory-backed models. We proposed using an extended TAM-based model to measure the relationship between perceived usefulness, perceived ease of use, quality of system, perceived enjoyment and intention to use constructs for a prototype research data sharing system based in blockchain technology. Although these constructs have been much investigated previously as antecedents to user acceptance of different technologies in various domains, this paper is the first to investigate the use of TAM for analyzing the factors influencing user acceptance of blockchain-based applications for sharing data, in this case research data among researchers. The value of this paper is mostly in the methodology, and it opens new directions to study distributed ledger technologies and decentralized applications from the user behavioral modeling perspective. We implemented the descriptive statistic, measurement models and structural models to present our results and used SEM analysis to observe the users' acceptance of the proposed blockchain-based system. In future work, we will investigate a new variable in the extended TAM, the *trust* that the user has in the blockchain-based system and we will perform the study in our actual system, rather than in a prototype with sketchy UI. We also hope to recruit a larger and broader participants pool in future studies.